\documentclass[wssquare]{ws-procs975x65}
\usepackage{graphicx}
\usepackage{wrapfig}

\newcommand\wlim{\mathop{\mbox{w--lim}}}
\newcommand{\liml}{\lim_{\lambda \to 0}}
\providecommand\arXiv[1]{\href{http://arXiv.org/abs/#1}{{\tt [\url{arXiv:#1}]}}}
\providecommand\cqg{Class.~Quant.~Grav.\ }

\providecommand\inlinecite[1]{Ref.~\citenum{#1}}

\usepackage{hyperref}
\hypersetup{
  colorlinks=true,
  citecolor=blue,
  linkcolor=magenta,
  filecolor=cyan,      
  urlcolor=cyan,
}

\makeatletter
\newcommand\sectionlessspacehack{
  \@startsection{section}{1}{\z@}{-10pt \@plus-3pt minus-6pt}{3pt}{\sectionfont}}
\newcommand\subsectionlessspacehack{
  \@startsection{subsection}{2}{\z@}{-10pt \@plus-3pt minus-6pt}{2pt}{\subsectionfont}}
\makeatother

\begin{document}
\hypersetup{pdfpagelabels=false,breaklinks=false,linktocpage}

\markboth{Ostrowski \& Roukema}
         {On the Green and Wald formalism}

\title{On the Green and Wald formalism}

\author{Jan J. Ostrowski$^{1,2,\dagger}$ and Boudewijn F. Roukema$^{1,2}$}

\address{$^1$Toru\'n Centre for Astronomy, 
  Faculty of Physics, Astronomy and Informatics,
  Grudziadzka 5,
  Nicolaus Copernicus University, ul. Gagarina 11, 87-100 Toru\'n
Poland\\
$^\dagger$E-mail: ostrowski~at~astro.uni.torun.pl}

\address{$^2$Universit\'e de Lyon, Observatoire de Lyon, Centre de
  Recherche Astrophysique de Lyon,\\ CNRS UMR 5574: Universit\'e
  Lyon~1 and \'Ecole Normale Sup\'erieure de Lyon,\\ 9 avenue Charles
  Andr\'e, F--69230 Saint--Genis--Laval, France\footnote{BFR: during
    invited lectureship; JJO: during long-term visit.}}

\begin{abstract} 
  Backreaction in the cosmological context is a longstanding problem
  that is especially important in the present era of precise
  cosmology. The standard model of a homogeneous background plus
  density perturbations is most probably oversimplified and is
  expected to fail to fully account for the near-future observations
  of sub-percent precision. From a theoretical point of view, the
  problem of backreaction is very complicated and deserves careful
  examination. Recently, Green and Wald claimed in a series of papers
  to have developed a formalism to properly describe the influence of
  density inhomogeneities on average properties of the Universe, i.e., the
  backreaction effect.  A brief discussion of this framework is
  presented, focussing on its drawbacks and on misconceptions
  that have arisen during the ``backreaction debate''.
\end{abstract}



\bodymatter

\sectionlessspacehack{A brief history of the ``backreaction debate''}
The subject of backreaction in cosmology has quite a long
history, given that modern cosmology itself is a young
discipline. In general, backreaction can be understood as the
influence of density inhomogeneities on average properties of the Universe
that are usually described by the Friedman--Lema\^{\i}tre--Robertson--Walker (FLRW) metric,
and in turn, their influence on light propagation and observables. Several approaches have
been developed to estimate the magnitude of backreaction, with a wide
range of results, depending on the method used. The
smallest effects are found in the perturbative regime, and the biggest 
for scalar averaging and for exact solutions. Disregarding
the issue of the magnitude of backreaction, most of the methods show that
this effect may act like dark energy,
i.e., produce an effective large-scale expansion in addition
to that of a strictly FLRW model.

Here, we focus on a particular framework proposed to
address the problem of backreaction, the Green and Wald
formalism, which reaches the opposite conclusions, i.e., that backreaction
is trace-free and cannot mimic a dark energy component. This claim
contradicts the results obtained by several different methods,
and thus became a subject of debate which we refer to here as the
``backreaction debate''.

Key ``backreaction debate'' articles, for ($+$) and against ($-$) backreaction {\em potentially} being able to mimic
dark energy, include the following:
\vspace{-1ex}
\begin{list}{$-$}{}
 \item \parskip=0.5ex
   Ishibashi and Wald: {\em``Can the acceleration of our
   Universe be explained by the effects of
   inhomogeneities?''} \cite{Ishibashi}---the authors claim
   that backreaction is negligible based on the smallness of
   the derivatives of metric deviations.
 \item 
  The Green and Wald formalism: {\em``New
  framework for analyzing the effects of small scale
  inhomogeneities in cosmology''} \cite{GW1}---the authors
  claim to have developed a formalism for a mathematically
  rigorous treatment of backreaction, with surprising
  results: the only effect that the density inhomogeneities
  can have on the background dynamics is via gravitational
  radiation.
\item 
  Examples by Green and Wald: {\em``Examples of backreaction of
  small-scale inhomogeneities in cosmology''} \cite{GW2}---the
  authors present two examples of their framework: a vacuum
  space--time, and a metric with an associated
  stress--energy tensor that violates the weak energy
  condition.
\item[$+$] {{Rebuttal by Buchert et al.: {\em``Is there
    proof that backreaction of inhomogeneities is irrelevant
    in cosmology?''}}} \cite{Buchert}---this paper presents a
  critical analysis of the Green and Wald formalism;
\item 
  Response to the rebuttal, by Green and Wald: {\em``Comments on
  Backreaction''} \cite{reply}---the authors reassert their main
  results from previous articles\footnote{Commented on in the final
    version of the rebuttal paper.}.
\end{list}        
\vspace{-1ex}
There is an important difference between the first and
second of these articles. In
\inlinecite{Ishibashi}, the authors use the arguments from 
standard perturbation theory, i.e., that not only does the metric of the real Universe
barely deviate from the FLRW metric, but its 
derivatives are also small. The same applies to 
stress--energy tensor perturbations, and the perturbed
Einstein equation reads 
\vspace{-1ex}
\begin{equation}
  \delta G_{ab} = 8\pi \delta T_{ab}\,. 
  \vspace{-1ex}
\end{equation}
This formulation has as an obvious limitation in
describing our Universe---we know that density
perturbations are much greater than one in amplitude at recent
epochs, on length scales well above those of ``black holes and neutron stars''
\cite{GW1}, implying---through the Einstein equation---the importance of curvature
(i.e., second derivatives of the metric). 
In addition,
as noted by Green and Wald in \inlinecite{GW1}, there is no
particular reason why the metric derivatives should also be
small. In \inlinecite{GW1}, the metric
first derivatives are allowed to be large but finite, and 
no constraints are placed on the second derivatives 
(in practice, however, for the formalism to give a non-zero result,
specific constraints have to be put on the second
derivatives \cite{Buchert}). Furthermore, no constraints are made on the
stress-energy tensor. All of these are desired features of a
good backreaction model; Green and Wald's effort to take
these into account deserves credit. Unfortunately, in our
opinion, the authors failed to 
provide a physically valid general statement, as we outline
below.

\sectionlessspacehack{The Green and Wald formalism}
 In this section we briefly describe the Green and Wald
 formalism (see \inlinecite{GW1} for 
 details). The formalism is based on the existence of a
 $\lambda$-dependent family of metrics $g_{ab}(\lambda,x)$
 on an arbitrary background manifold ($M$) that are close
 to a background metric $g_{ab}(0,x)$, and which converge to the latter
 point-wise as the scalar parameter $\lambda$ approaches 
 zero. This assumption is followed by another four
 concerning the behaviour of the metric and its derivatives:
\begin{romanlist}[(ii)]
\item \parskip=1ex
  For all $\lambda > 0$ the metric $ g_{ab}(\lambda,x)$
  satisfies: 
  \begin{equation}
G_{ab}(g(\lambda,x))+ \Lambda
  g_{ab}(\lambda,x) = 8\pi T_{ab}(\lambda),
  \label{eq-GW-assump-i}
  \end{equation} where
  $T_{ab}(\lambda)$ obeys the weak energy condition.
\item 
  There exists a smooth function $C_1(x)$ on $(M,g(0))$ such that:
\begin{equation}
|h_{ab}(\lambda,x)| \leq \lambda C_1 (x) \;\; ; \;\; h_{ab}(\lambda,x) = g_{ab}(\lambda,x) - g_{ab}(0,x) .
  \label{eq-GW-assump-ii}
\end{equation} 
\item 
  There exists a smooth function $C_2(x)$ on $(M,g(0))$ such that
  $|\nabla_c  h_{ab}(\lambda,x)| \leq C_2 (x),$ i.e. 
  the derivatives do not have to be small.
\item 
  There exists a smooth tensor field $\mu_{abcdef}$ on $(M,g(0))$
  such that\footnote{Eq.~(9) of Buchert et al. \protect\cite{Buchert} has a typo here: $|\cdot|$ should
    read $[\cdot]$.}:
  \begin{equation}
    \wlim_{\lambda \searrow 0}[\nabla_a
  h_{cd}(\lambda,x) \nabla_b h_{ef}(\lambda,x)] =
  \mu_{abcdef}.
  \label{eq-GW-assump-iv}
  \end{equation}
\end{romanlist}
The last assumption uses the notion of the weak limit,
$\wlim$: we say that an arbitrary tensor field $A_{a_1
  \ldots a_n}(\lambda)$ converges weakly to $B_{a_1 \ldots a_n}$,
i.e., $\wlim_{\lambda \searrow 0} A_{a_1 \ldots a_n}(\lambda) =
B_{a_1 \ldots a_n} $, if
$\forall\,f^{a_1  \ldots a_n}$ of compact support,
\begin{equation}
 \lim_{\lambda \searrow 0}\int f^{a_1\ldots a_n} A_{a_1
  \ldots a_n}(\lambda) = \int f^{a_1\ldots a_n} B_{a_1 \ldots a_n}.
\end{equation}
The integration is performed over a 4D region of space--time
fixed in the background manifold.  With these five
assumptions, the Einstein equation for the
background metric is derived by comparing and manipulating the
curvature terms. The Green and Wald equations for the
background metric $g_{ab}(0,x)$ then read:
\begin{equation}
\wlim_{\lambda \searrow 0} \left[
G_{ab}\left(g_{ab}(0,x)\right) \right] + \Lambda g_{ab}(0,x) = 8\pi
 \wlim_{\lambda \searrow 0}\left[T_{ab}(\lambda)+t_{ab}(\lambda)
 \right],
\label{e-GW-eqns-wlim}
\end{equation} 
where: 
\begin{equation}
 t_{ab}(\lambda) =
 2\nabla_{[a}C^e_{\phantom{e}e]b}-2C^f_{\phantom{f}b[a}C^e_{\phantom{e}e]f}
 -g_{ab}(\lambda)g^{cd}(\lambda)\nabla_{[c}C^e_{\phantom{e}e]d} 
 +g_{ab}(\lambda)g^{cd}(\lambda)C^f_{\phantom{f}d[c}C^e_{\phantom{e}e]f}
 ,
\end{equation}
and  
\begin{equation}
\nonumber C^c_{\phantom{c}ab} =
\frac{1}{2}g^{cd}(\lambda)\left[\nabla_ag_{bd}(\lambda) +
  \nabla_bg_{ad}(\lambda) - \nabla_dg_{ab}(\lambda)\right].
\end{equation}
We can now give names to the terms in (\ref{e-GW-eqns-wlim}):
\begin{equation}
 G_{ab}(g^{(0)}) + \Lambda g^{(0)}_{ab} = 8\pi T^{(0)}_{ab}+8\pi t^{(0)}_{ab} ,
\end{equation} 
where according to Green and Wald, $T_{ab}^{(0)}$, defined
\begin{equation}
 \wlim_{\lambda \searrow 0}T_{ab}(\lambda) = T^{(0)}_{ab} , 
\end{equation}
represents a stress--energy tensor averaged over small-scale
inhomogeneities.  
Green and Wald examined an ``effective'' stress-energy tensor $t^{(0)}_{ab}$, inferring that:
\begin{itemlist}
\item 
  $t^{(0)}_{ab}$ is trace-less, i.e. ${t^{(0)a}}_{a} =0$ ; \label{t1}
\item 
  $t^{(0)}_{ab}$ obeys the weak energy condition
  i.e. $t^{(0)}_{ab}t^at^b \geq 0$ \label{t2}.
\end{itemlist}
\vspace{-0.5ex}
To put it in words: $t^{(0)}_{ab}$ cannot mimic dark energy. 

\sectionlessspacehack{Discussion}
This eventually led to the rebuttal paper by Buchert et. al
\cite{Buchert}. Shortly after this appeared on the {\em ArXiv}
preprint server, Green and Wald published a response in which they
uphold their previous statements. Additionally, they clarify
the domain of applicability of their formalism in
relation to popular approaches to backreaction. In particular, 
Green and Wald state that their formalism does not apply to situations when:
\begin{itemize}
\item 
  the actual metric (e.g., at recent epochs) is far from FLRW; or
\item 
   one wishes to construct an effective metric (or other effective
   quantities) through some averaging procedure
\end{itemize}    

This, in principle, ends the debate about whether backreaction 
has been excluded as a dark energy candidate: the Green and Wald formalism
does not apply to the main body of backreaction research; backreaction 
remains a viable dark energy candidate.
We briefly outline some characteristics that the formalism lacks.

\subsectionlessspacehack{Backreaction without backreaction} 
Let us introduce a more precise definition of backreaction
(setting $\Lambda=0$ for simplicity). Assume
that we know the inhomogeneous metric describing the real Universe and
that we want to derive its averaged dynamical behaviour on a certain scale. We
do this by applying a procedure such that by smoothing over larger
and larger scales of inhomogeneities, we end up with a background metric (on 
a scale that we accept as homogeneous). However, to construct the Einstein tensor, and
thus to describe the averaged dynamics, we need not only the metric
but also the metric derivatives and products of derivatives. In general,
averaging and differentiating or taking products of derivatives are
non-commuting operations \cite{EllisBuch05}.
Thus, while it is trivially true that, provided 
that an averaging procedure $\langle . \rangle$ is properly defined,
 $ \langle G_{ab}(g_{ab}) \rangle = 8\pi \langle T_{ab} \rangle$,
we cannot expect that
\begin{equation}
  \label{noback}
  G_{ab}(\langle g_{ab} \rangle) = 8\pi \langle T_{ab} \rangle.
\end{equation}
Backreaction
($\tau_{ab}$) is then the term compensating the discrepancies coming from
this non-commutativity:
\begin{equation}
  G_{ab}(\langle g_{ab} \rangle) = 8\pi \left(\langle T_{ab} \rangle +
  \tau_{ab}\right) .
  \label{e-tau-backreaction}
\end{equation} 
This term should, in principle, be present at each intermediate scale
between a small scale and the homogeneous scale.

Let us take a look at 
(\ref{eq-GW-assump-i}), i.e.
Green and Wald's assumption (i), which becomes
\begin{equation}
  G_{ab}(g(\lambda,x)) = 8\pi T_{ab}(\lambda),
  \label{eq-GW-assump-i-Lambda-zero}
\end{equation}
which we can treat, from the lack of any other options, as a
definition of the $\lambda$-dependent family of stress--energy
tensors.  Suppose that for some $\lambda > 0$, $T_{ab}(\lambda)$ is
smoother than the inhomogeneous, unsmoothed stress-energy tensor
$T_{ab}(1)$, thanks to an averaging procedure.  Then a backreaction
term $\tau_{ab}(\lambda)$ must appear in the $\lambda$-dependent
Einstein equation, due to the non-commutativity of averaging
(\ref{e-tau-backreaction}).  But this contradicts
(\ref{eq-GW-assump-i-Lambda-zero}).  Hence, in the Green and Wald
formalism, the only averaging allowed for $\lambda > 0$ is exactly
commutative averaging ($\tau_{ab}(\lambda) = 0 \; \forall \lambda >0$),
and a backreaction term $t_{ab}(0)$ suddenly blips on at $\lambda=0$,
discontinuously \cite{Buchert}.  Thus, it is difficult to find a physically
meaningful interpretation of this limiting procedure.

\subsectionlessspacehack{Averaging without averaging}
There is no overall agreement on a physically meaningful way to average the Einstein equation. One can quite quickly run into conceptual problems trying
to perform averaging of fields on manifolds in a unique and
gauge-independent way. In their formalism, Green and Wald 
bypass these difficulties by being very general, and yet they are able to extract some crucial features of averaged equations. The role of the averaging operator is played by the weak limit specified up to some arbitrary well-behaving test tensor field used to contract the averaged quantities. According to Green and Wald, the action of the weak limit can be interpreted as follows
(\cite{GW1}): ``{\it{Roughly speaking, the weak limit performs a local
    space-time average of $A_{a_1 ... a_n}(\lambda)$ before letting
    $\lambda \rightarrow 0$}}'', where $A_{a_1 ... a_n}(\lambda)$ is a
one-parameter family of tensor fields. For this to be non-trivial, the limit
operator and the integral cannot commute: $\liml \int \neq \int \liml$,
since otherwise, the averaging would be performed on the background value of the tensor field and the integration would become redundant for the homogeneous
background. 

\begin{wrapfigure}[12]{r}{0.5\textwidth}
  \centering
  \vspace{-8ex}
  \includegraphics[width=0.45\columnwidth]{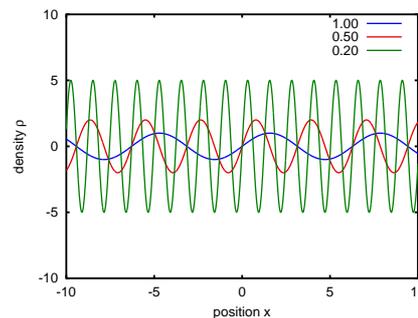}
  \vspace{-1.5ex}
  \caption{Schematic illustration of density $\rho := \lambda^{-1}$
    $\sin(x/\lambda) = \mathrm{d}^2/\mathrm{d}x^2\, h$, where 
    $h :=-\lambda \sin (x/\lambda).$
    As $\lambda \searrow 0$, $h$
    approaches zero pointwise, but $\rho$ is pointwise unbounded.
    \label{f1}}
  \vspace{-3ex}
\end{wrapfigure}

However, for a fixed domain of integration, 
fulfilling this requirement of non-commutativity 
requires breaking one of the dominant
convergence theorem assumptions (e.g., Figure~\ref{f1}). 
Indeed, in the examples provided by Green and Wald, the derivatives and
the products of derivatives, while being continuous in the space--time
coordinates, do not have pointwise limits as $\lambda \searrow 0$.


Figure~\ref{f1} shows the behaviour of a toy model 
$\rho(x) = \mathrm{d}^2/\mathrm{d}x^2\, h(x)$ (analogous to the usual density
$\rho$ that relates to the second derivatives of the metric via the
Einstein equation),
where $h$
approaches zero pointwise, while $\rho$ is $\lambda$-discontinuous
(it does not have a pointwise limit).
Instead of successive coarse-graining, the
density profile approaches homogeneity by a {\em fine}-graining process,
which is the opposite of what we would expect from an averaging
procedure.
(In contrast, GW's procedure {\em does} have 
features likely to be relevant to gravitational radiation.)

Moreover, integrating by parts, we have (e.g. for $A_{a_1...a_n} := h_{ab}$):
\vspace{-0.5ex}
\begin{align}
\liml \int & (\nabla_b A_{a_1...a_n}(\lambda))f^{b,a_1...a_n} =
\nonumber \\
 &\liml A_{a_1...a_n}(\lambda)f^{b,a_1...a_n} - \int \liml  A_{a_1...a_n}(\lambda)\nabla_b f^{b,a_1...a_n} = 0.
\vspace{-1ex}
\end{align}
Thus, the apparently non-local integration of derivatives in GW's procedure 
reduces to pointwise operations; it misses generic features of averaging: non-locality and scale-dependence.

\sectionlessspacehack{Conclusions}
The Green
and Wald formalism is not mathematically general---see
\inlinecite{Buchert} for the hidden assumptions---and its physical interpretation is
far from obvious. An unaware reader might be tempted to think that
mathematically general arguments against common approaches to
backreaction (e.g., exact inhomogeneous cosmological solutions or the
Buchert equations) were presented, but this is not the case.

\section*{Acknowledgments}
This contribution owes much to collaboration with T.~Buchert, M.~Carfora,
G.F.R.~Ellis, E.W.~Kolb, M.~MacCallum, S.~R\"as\"anen, L.~Andersson, 
A.~Coley \& D.~Wiltshire.
Part of this project was performed under grant 2014/13/B/ST9/00845 of the
National Science Centre, Poland. 









\end{document}